\documentstyle[12pt]{article}
\begin{document}
\begin{center}
{\Large\bf Vibrational Superposition States Without Rotating Wave 
Approximation}
\end{center}
\begin{center}
S. MANCINI$^{1,3}$, H. MOYA-CESSA$^{2,3,4}$ and P. TOMBESI$^{2,3}$
\end{center}
\begin{center}
$^1$ Dipartimento di Fisica, Universit\`a di Milano, 
Via Celoria 16, I-20133 Milano, Italy \\
$^{2}$ Dipartimento di Matematica e Fisica, Universit\`a di 
Camerino, I-62032 Camerino, Italy \\
$^3$ Istituto Nazionale per la Fisica della Materia, Unit\`a  di Camerino, Italy \\
$^4$ INAOE, Coordinaci\'on de Optica, Apdo. Postal 51 y 216, 72000 
Puebla, Pue., Mexico
\end{center}
\begin{abstract}
We propose a scheme to generate superpositions
of coherent states for the vibrational
motion of a laser cooled trapped-ion. It is based on the interaction with 
a standing wave making use of the counter-rotating terms, i.e.
not applying the rotating wave approximation.
We also show that the same scheme can be exploited
for quantum state measurement, i.e. with the same scheme non-classical 
states may be reconstructed.
\end{abstract}
\newpage
\section{Introduction}

In recent years there has been great interest in
quantum state preparation and measurement \cite{jmo}.
In particular the generation of non-classical 
states was proposed for light field \cite{optcat}
and for vibrational motion of a trapped ion  \cite{vogel}.
Consequentely, several 
schemes to reconstruct such states
have been developed \cite{ulf,milb,Cirac}.
In both cases (field and ion) key features are shared
for their generation and reconstruction, e.g. the 
relation of the atomic inversion with
quasiprobability distributions as they both
can be expressed as a sum of diagonal
density matrix elements (see for instance
\cite{Davidov,Moya}).
On the other hand, both systems,
atom-field and ion-laser interactions can
be described by the same kind of
Hamiltonians under certain conditions (Lamb-
Dicke or intensity approximations \cite{vogel}). 
Here we will concentrate 
on the vibrational motion of an
ion, bearing in mind that,
because of the similarity of such a system with the
atom-field interaction,
the scheme could be extended to the later case.
It was recently shown by Wilkens and Meystre
\cite{Wilk} that the use of the Rotating Wave Approximation
(RWA) (in the atom-field interaction case) does
not permit the possibility of many absorption
and/or emission processes, and
therefore it 
is needed a second mode of the electrmagnetic
field to change the situation: in
such a case,
although an absorption process is followed by
an emission process, there exist
situations in which  absorption from the first
mode may be followed by emission
into the 
second mode, then again by
absorption from the first mode, etc. This allows
coherences between distant
Fock states to take place.  
Here we show that
by not making use of the RWA,
the ion effectively becomes sensitive to
coherences between
such distant Fock states
giving rise to Schr\"odinger cat-like states \cite{cat}.
Moreover, we show that the same mechanism allows
the reconstruction of vibrational state.

\section{The model}

Let us consider  a trapped ion laser
cooled into the Lamb-Dicke limit using a
strong transition and located at the node
of an optical standing wave (SW). Such
system can be described by
the Hamiltonian \cite{milb} (in a frame rotating at
the SW frequency $\omega_L/2\pi$)
\begin{equation}
\hat{H}= \hbar \nu \hat{n}  
+ \hbar\Delta\hat{\sigma}_z +\hbar\eta\Omega  
\hat{X} \hat{\sigma}_{x},
\label{1}
\end{equation}
where $\nu$ is the oscillation angular frequency of the ion in the trap, 
$\Delta=\omega_0-\omega_L$, is the detuning between  the atomic 
($\omega_0$)  and the SW angular frequencies, 
$\Omega$ is the Rabi frequency for the
two-level transition, $\eta$ is the Lamb-
Dicke parameter ($\eta<<1$).  
$\hat{n}=\hat{a}^{\dagger}\hat{a}$ is the number operator and $\hat{X}= 
\hat{a}^{\dagger}+\hat{a}$ is  
the generalized position operator, 
with 
$\hat{a}^{\dagger}$ and $\hat{a}$ the creation and annihilation operators 
respectively for the vibrational mode.
$\hat{\sigma}_z$ is the inversion operator
and $\hat{\sigma}_{x} =\hat{\sigma}_{+}+\hat{\sigma}_{-}$ is the polarization 
operator, where 
$\hat{\sigma}_{+}$ and $ \hat{\sigma}_{-}$ are the atom raising and lowering 
operators, respectively.
By considering the resonant case, i.e. $\Delta=0$,
from the Hamiltonian (\ref{1}) we can
obtain the evolution operator in the following form
(see for instance \cite{manci})
\begin{equation}
\hat{U}(\tau)=e^{i\phi (\tau)} e^{-i\hat{n}\tau} 
\hat{D}[\alpha(\tau)\hat{\sigma}_x],
\quad
\tau=\nu t\,,
\label{2}
\end{equation}
where we have defined
\begin{equation}
\hat{D}[\alpha(\tau)\hat{\sigma}_x] = 
\exp\left\{\hat{\sigma}_x[\alpha(\tau)\hat{a}^\dagger-
\alpha^*(\tau)\hat{a}]\right\}\,,
\label{3}
\end{equation}
and
\begin{equation}
\phi (\tau)= \eta^2\frac{\Omega^2}{\nu^2}(\tau - \sin\tau), \ \ \ \ \ 
\alpha (\tau) =\eta \frac{\Omega}{\nu}(1-e^{i\tau}).
\label{4}
\end{equation}
Let us now consider the generic initial state
\begin{equation}\label{ini}
|\Psi(0)\rangle=|\psi_e\rangle|\psi_v\rangle\,,
\end{equation}
where the atomic initial state is
\begin{equation}
|\psi_e\rangle={\cal A}|g\rangle+{\cal B}e^{i\varphi}|e\rangle\,,
\end{equation}
with $\varphi$, ${\cal A}$, and ${\cal B}$ real numbers
such that ${\cal A}^2+{\cal B}^2=1$; $|e\rangle$ and $|g\rangle$
represent two different states of the atom.
Then, the state (\ref{ini}) evolves according to
\begin{eqnarray}
|\Psi(\tau)\rangle=e^{i\phi(\tau)}
e^{-i{\hat n}\tau}
\Bigg\{{\cal A}\left[
\cosh\left(\alpha(\tau){\hat a}^{\dag}-\alpha^*(\tau){\hat a}\right)
|g\rangle
+\sinh\left(\alpha(\tau){\hat a}^{\dag}-\alpha^*(\tau){\hat a}\right)
|e\rangle\right]\nonumber \\
+e^{i\varphi}{\cal B}
\left[
\sinh\left(\alpha(\tau){\hat a}^{\dag}-\alpha^*(\tau){\hat a}\right)
|g\rangle
+\cosh\left(\alpha(\tau){\hat a}^{\dag}-\alpha^*(\tau){\hat a}\right)
|e\rangle\right]
\Bigg\}|\psi_v\rangle\,.
\label{evol}
\end{eqnarray}

\section{Generation of superpositions of coherent states}

To this end, we consider the ion 
initially to be not excited, $|\psi_e\rangle=|g\rangle$,
and the vibrational motional state to be a
coherent state $|\psi_v\rangle=|\alpha_0\rangle$.
From Eq.(\ref{evol}), we easily get the evolved state.
For particular times, $\tau=(2q+1)\pi$ with $q$ integer number, 
we obtain
\begin{equation}
|\Psi(\tau)\rangle=\frac{1}{2}\Bigg\{
\Big[|-(\alpha_0+\alpha)\rangle+|\alpha-\alpha_0\rangle\Big]
|g\rangle
+\Big[|-(\alpha_0+\alpha)\rangle-|\alpha-\alpha_0\rangle\Big]
|e\rangle
\Bigg\}\,,
\end{equation}
where $\alpha=-2\eta\Omega/\nu$.
We have neglected the overall phase $\phi$
since it is not relevant.
It is now immediate to see that an atomic selective  measurement 
\cite{Garra}
yields a superposition of two
distinct coherent states
by means of the wave function collapse.
In particular, for $\alpha_0=0$,
it is possible to get the even or odd coherent states \cite{vim}
depending on the result $g$ or $e$ of the measurement. 
Furthermore, if we allow for more than
one interaction, superpositions of coherent
states in a line can be produced, i.e. any non-classical state
can be produced \cite{wal}.

\section{Vibrational State Measurement}

Let us now consider the reconstruction method
within the same scheme.
In this case the state $|\psi_v\rangle$ would be unknown.
To be more general, instead of a pure state $|\psi_v\rangle$,
we are going to consider 
an initial vibrational state ${\hat\rho}_v$. 
Under such assumptions, we can calculate
the probability of the ion being in the
ground state again from
Eq. (\ref{evol}), which can be written as
\begin{equation}\label{Pg}
P_g = \frac{1}{2}+
\frac{{\cal A}^2-{\cal B}^2}{2}{\rm Tr}_v
\left\{\cos\left(k{\hat X}_{\theta}\right) {\hat\rho}_v\right\}
-{\cal A}{\cal B}\sin\varphi{\rm Tr}_v
\left\{\sin\left(k{\hat X}_{\theta}\right) {\hat\rho}_v\right\}
\label{pg}
\end{equation}
where we have introduced the quadrature
\begin{equation}
{\hat X}_{\theta}={\hat a}e^{-i\theta}+{\hat a}^{\dag}e^{i\theta}
\end{equation}
and the radial and angular variables
\begin{equation}
k=2|\alpha(\tau)|=4\eta\frac{\Omega}{\nu}\sin\left(\frac{\tau}{2}\right)
\,,\quad
\theta=\arctan\left[\frac{\sin\tau}{1-\cos\tau}\right]-\frac{\pi}{2}
\end{equation}
Eq. (\ref{pg}) shows that the probability of the ion being in its
ground state is
proportional to the characteristic function \cite{Milbu} of the vibrational
motion of the ion
\begin{equation}\label{8}
\chi(k,\theta)={\rm Tr}_v\left\{e^{ik{\hat X}_{\theta}}{\hat \rho}_v\right\}
\end{equation}
with the property
$\chi(k,\theta+\pi)=\chi(-k,\pi)$.
This function contains all information about the state
\cite{Milbu,Sten} which
makes  Eq. (\ref{Pg}) a significant result.
It can also be related to the Shirley (or ambiguity)
function which is the totally off-diagonal complement of the
Wigner function \cite{Sten}.
Hence, by appropriately adjusting $\varphi$, ${\cal A}$ and ${\cal B}$
it is possible to measure both the even and the odd part of the
characteristic function.

For what concerns the argument $k$, it can be varied by changing the
Rabi frequency $\Omega$, i.e. by changing the intensity of the standing
wave. Instead, the quadrature phase $\theta$, can be varied by means of the
interaction time $\tau$; but because the latter affects also $k$,
it is preferable to set the desired quadrature phase $\theta$
with a free evolution
prior the interaction,
keeping $\tau$ equal for all sets of measurements.
Moreover, there exist standard simple techniques to
monitor the probability of the ion being in its ground
(excited) state (see \cite{milb,Cirac}).
The measurement of $\chi$ allows in reality the direct
sampling of density
matrix elements in some representations
since we can write
\begin{equation}
{\hat \rho}_v=\int_{-\infty}^{\infty}
\int_0^{\pi}\frac{d\theta}{\pi}\chi(k,\theta)\,K({\hat X_{\theta}})
\,,\quad\quad
K({\hat X_{\theta}})=\frac{|k|}{4}\exp\left[-ik{\hat X}_{\theta}\right]\,,
\end{equation}
and the kernel operator $K$ results bounded e.g. in number representation
\cite{cagla}.
Instead, direct sampling of the Wigner function is not possible
since the corresponding kernel is not bounded.
The analogous problem exist in optical homodyne
tomography, where the measured quantity
is the Fourier transform of the characteristic function \cite{dar}.

\section{conclusions}

We have considered a trapped ion
interacting with a radiation field, and
we have shown that
by exploiting counter-rotating terms
it is possible to produce and retrieve
quantum coherences in the vibrational degree of freedom.
The significance of our method relies on its simplicity,
as we do not need to do any further assumptions but the standard one of
small Lamb-Dicke parameter.

{\large\bf Acknowledgments}

This work was partially supported by
Consejo Nacional de Ciencia y Tecnolog\'ia
(CONACyT), Mexico.

%
%
%
%

\begin{thebibliography}{xxxx}
\bibitem{jmo}
{\it J. Mod. Opt.}, 1997, {\bf 44}, N.11/12, Special Issue: Quantum State
Preparation and Measurement.
\bibitem{optcat}
{\sc Yurke, B.,} and {\sc Stoler, D.}, 1986, {\it Phys. Rev. Lett.},
{\bf 57}, 13; 1992, {\sc  Brune, M.,  Haroche, S., 
Raimond, J.M., Davidovich, L.,} and {\sc Zagury, N.,} {\it  Phys. Rev.} A, 
{\bf 45}, 5193.
\bibitem{vogel}
{\sc de Matos Filho, R.L.,}  and {\sc  Vogel, W.,} 1996, {\it Phys. Rev. Lett.,} 
{\bf 76}, 608; 1996, {\it Phys. Rev.} A, {\bf 54}, 4560.
\bibitem{ulf}
{\sc Leonhardt, U.,}, 1997, {\it Measuring  the Quantum State of Light} (Cambridge: CUP).
\bibitem{milb}
{\sc D'Helon, C., and {\sc Milburn, G.J.,}1996, {\it Phys. Rev.} A, 
{\bf 54}, R25.
\bibitem{Cirac}
{\sc Cirac,, J.I., Blatt, R., Parkins, A.S.,} and {\sc Zoller, P.,} 1994,
{\it Phys. Rev.} A, {\bf 49}, 1202.
\bibitem{Davidov} 
{\sc Lutterbach, L.G.,} and {\sc Davidovich, L.,} 1997, {\em Phys. Rev. Lett.} {\bf 78}, 2547.
\bibitem{Moya}
{\sc Moya-Cessa, H., Dutra, S.M., Roversi}, J.A.,}  and  {\sc Vidiella-Barranco, A.,}
{\em J. Mod. Opt.}, 1999, {\bf 46}, 555; 
{\sc Moya-Cessa, H., Roversi, J.A., Dutra, S.M.,}   and  {\sc Vidiella-Barranco, A.,}
{\em Phys. Rev.} A, 1999, {\bf 60}, 4029.
\bibitem{Wilk}
{\sc  Wilkens, M.}, and {\sc Meystre, P.,} 1991,
{\it Phys. Rev.} A, {\bf 43}, 3832;
see also  {\sc  Dutra, S.M.,  Knight, P.L.,} and {\sc Moya-Cessa, H.,} 1993,
{\it Phys. Rev.} A., {\bf 48}, 3168; {\sc Kim, M.S.} and {\sc Agarwal, G.S.,} 1999,
{\it Phys. Rev.}, A {\bf 59}, 3044.
\bibitem{cat}
{\sc Schr\"oedinger, E.,} 1935, {\it Naturwissenschaft.}, {\bf 23}, 807,; {\it ibid} 823;
{\it ibid} 844.
\bibitem{manci}
{\sc Mancini, S.,  Man'ko, V.I.,} and {\sc Tombesi, P.,} 1997,
{\it Phys. Rev.} A, {\bf 55}, 3042;
{\sc Bose, S., Jacobs, K.,} and {\sc Knight, P.L.,} 1997,
{\it Phys. Rev.} A, {\bf 56}, 4175.
\bibitem{Garra}
{\sc Garraway, B.M., Sherman, B., Moya-Cessa, H., Knight, P.L.,} and {\sc Kurizki, G.,}
1994, {\it Phys. Rev.} A, {\bf 49}, 535.
\bibitem{vim}
{\sc Dodonov, V.V.,  Malkin, I.A.,} and {\sc Man'ko, V.I.}, 1974,
{\it Physica}, {\bf 72} , 597.
\bibitem{wal}
{\sc Janszky, J., Domokos, P.,} and {\sc Adam, P.,} 1993,
{\it Phys. Rev.} A, {\bf 48}, 2213; 
{\sc Moya-Cessa, H., Wallentowitz, S.,} and {\sc Vogel, W.,} 1999,
{\it Phys. Rev.} A, {\bf 59}, 2920.
\bibitem{Milbu} 
{\sc Walls, D.F.,} and {\sc Milburn, G.J.,} 1994, {\it Quantum Optics} (Springer, Berlin).
\bibitem{Sten} 
{\sc Stenholm, S.,} and {\sc Vitanov, N.,} 1999, {\it J. Mod. Opt.}, {\bf 46}, 239.
\bibitem{cagla}
{\sc  Cahill, K.E.,} and {\sc  Glauber, R.J.,} 1969,
{\it Phys. Rev.}, {\bf 177}, 1857.
\bibitem{dar}
{\sc D'Ariano, G.M.,} 1997, in {\it Quantum Optics and Spectroscopy of Solids},
T. Hakioglu and S. Shumovsky Eds. (Kluwer, Amsterdam), p.175.
\end{thebibliography}
\end{document}